\documentclass[slac_one]{revtex4}
\usepackage{graphicx}
\usepackage{fancyhdr}
\begin{document}

\title{Search for Dark Matter} 

%

\author{Graciela B. Gelmini}
\affiliation{Department of Physics and Astronomy, UCLA, 475 Portola Plaza, Los Angeles, CA 90095, USA}

\begin{abstract}
The search for dark matter is a very wide an active field of research, and I necessarily concentrate here only in some aspects of it. I will review the prospects for direct and indirect dark matter searches  of Weakly Interacting Massive Particles in the dark halo of our galaxy and focus in particular on the data of GLAST, PAMELA and DAMA. 
 \end{abstract}

\maketitle

\thispagestyle{fancy}


\section{WHAT WE KNOW ABOUT DARK MATTER} 
We know a lot about dark matter (DM). First of all we know it exists: modified gravity models cannot explain the spatial segregation between the regions where most of the mass is and where most of the visible matter is  in the ``bullet cluster"~\cite{bullet-cluster} (a cluster formed by two colliding galaxies). In this cluster, the regions where
most of the gravitational potential is (as mapped though gravitational lensing) and where most of the visible matter  (the X-ray emitting cluster of baryonic matter) is, are separated. Any modified gravity model predicts instead a larger gravitational field where the visible matter is. We know the abundance of the DM in the Universe, about 23\% of the total density, at the few percent level: the Cosmic Microwave Background (CMB) anisotropy data from 5 years of observation by WMAP, combined with distance measurements from the Type Ia supernovae (SN) and the Baryon Acoustic Oscillations (BAO) give 
$\Omega_{\rm CDM} = 0.233 \pm 0.013$~\cite{WMAP-5year}. We know most of the DM in the dark galactic halo of our galaxy does not consist of MACHOs (Macroscopic Astrophysical Compact Halo Objects) with masses  larger than 10$^{-7}$ 
of the solar mass~\cite{MACHOs}.  We also know that most of the DM is not baryonic, since CMB, SN and BAO data, in agreements with Big-Bang Nucleosynthesis (BBN) measurements, imply that the abundance of baryons is only 4-5\% of the total density: $\Omega_{\rm baryons}=0.0462 \pm 0.0015$~\cite{WMAP-5year}.

We know also that the  DM cannot be explained within the Standard Model (SM) of elementary particles. Among the particles in the SM only neutrinos constitute part of the DM, but they are Hot DM (HDM) and the bulk of the DM can only be either Cold (CDM) or Warm (WDM). HDM particles are relativistic at the moment galaxies should start forming in the early Universe, at temperatures $T \simeq$ keV. WDM particles, instead, are semi-relativistic and CDM particles are non-relativistic at those temperatures. We know HDM cannot constitute more than 0.02 of the DM.  There are no WDM or CDM candidates in the Standard Model, but there are many in all extensions of the SM. For example, sterile neutrinos, gravitinos and non-thermal neutralinos can be WDM. WIMPs, Weakly Interacting Massive Particles (such as the  LSP, the Lightest Supersymmetric Particle,  or variants of it:  LKP- the Lightests Kaluza-Klein particle, LZP-  the Lightest Z3-charged Particle in warped SO(10) models with a Z3 discrete symmetry, LTP- the Lightest T-odd heavy Photon in the Little Higgs model with T-parity...), axions, WIMPZILLAS,  or solitons (Q-balls) could be CDM.  

Because of spontaneous symmetry breaking arguments totally independent of the DM issue, we do expect new physics beyond the SM to appear at the electroweak scale soon to be explored by the Large Hadron Collider (LHC) at CERN.  Naturalness arguments imply that above the TeV scale there should be cancellations in the radiative corrections to the SM Higgs mass due to a new theory, such as supersymmetry, SUSY, with or without composite Higgs bosons,  technicolor, either walking or top-assisted, larger extra spatial, possibly warped, dimensions, or the Little Higgs model, in which the Higgs is a 
pseudo-Goldstone  boson.  All of these extensions of the SM provide the main potential discoveries at the LHC and also DM candidates.  Thus, physics beyond the SM is required by the DM and expected at the electroweak scale, but both new physics may or may not be related! Thus the experiments at the LHC and the searches for the DM in our galactic halo are independent and complementary. We will concentrate in the following on WIMP DM searches, which are both complementary to LHC searches~\cite{LHC-DM-ICHEP08} and to each other.
 
 Direct DM detection searches look for energy deposited within a detector by the collisions  of  WIMPs belonging to the dark halo of our galaxy. The DM signature we expect to observe in these detectors is an annual modulation of the signal and the same  WIMP mass and cross section   seen by different experiments using different target nuclei. I will in particular mention the
 DAMA modulation signal.
 
 Indirect DM searches look for WIMP annihilation (or sometimes decay) products, be them neutrinos from the center of the Sun or the Earth (searched for by AMANDA, IceCUBE~\cite{Rott} or the future KM3NeT~\cite{deJong} in the Mediterranean) or anomalous cosmic rays, such as positrons and antiprotons,   or $\gamma$ rays from the galactic halo  or the galactic center  (observed from space by GLAST (now FGST)~\cite{GLAST-ICHEP08}, PAMELA~\cite{Boezio} or AMS~\cite{Romana}, from ground observatories such as HESS, VERITAS, CANGAROO or MAGIC, or balloon experiments such as HEAT). The DM signature here would be a signal that cannot be reproduced by astrophysical sources or cosmic rays, and possibly  confirmed by more than one type of experiment. I will concentrate here  in the physics potential of GLAST and PAMELA.
  
 \section{INDIRECT DARK MATTER SEARCHES} 
 
 WIMP candidates $\chi$ are usually neutral Majorana fermions  or bosons, i.e. particle and antiparticle are identical and can annihilate in pairs.  The maximum energy of the annihilation products is the WIMP mass $m_\chi$, because WIMPs in the galactic halo are non-relativistic at present.   Photons are particularly interesting annihilation products because they point back to their source (below 10's of TeV of energy).  The annihilation signal depends on the square of the WIMP density, thus the best photon sources are regions of high WIMP density: the galactic center of our galaxy or of nearby galaxies and DM clamps,  or subhaloes, remaining as substructure within our galactic halo. Galactic haloes grow hierarchically, incorporating lumps and tidal streams from earlier phases of structure formation. The existence of clumps of higher density boost  the annihilation signal (by a boost factor $B$) with respect to what would be expected from an homogeneous halo. The value of $B$ is not known and large $B$ values are usually needed for WIMP annihilation signals to be observable~\cite{subhaloes}. 
 
\begin{figure*}[t]
\includegraphics[width=70mm]{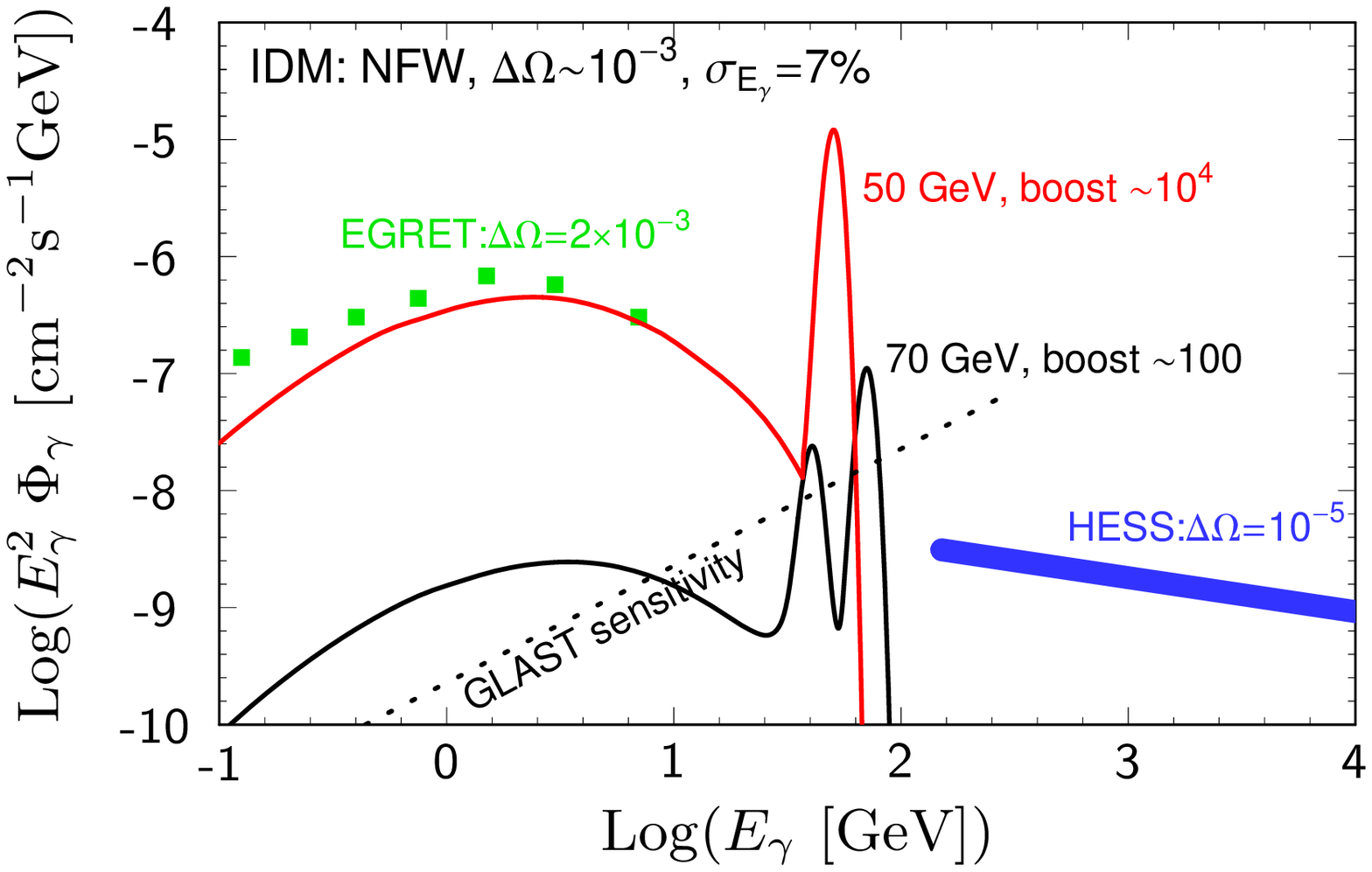}
\includegraphics[width=72mm]{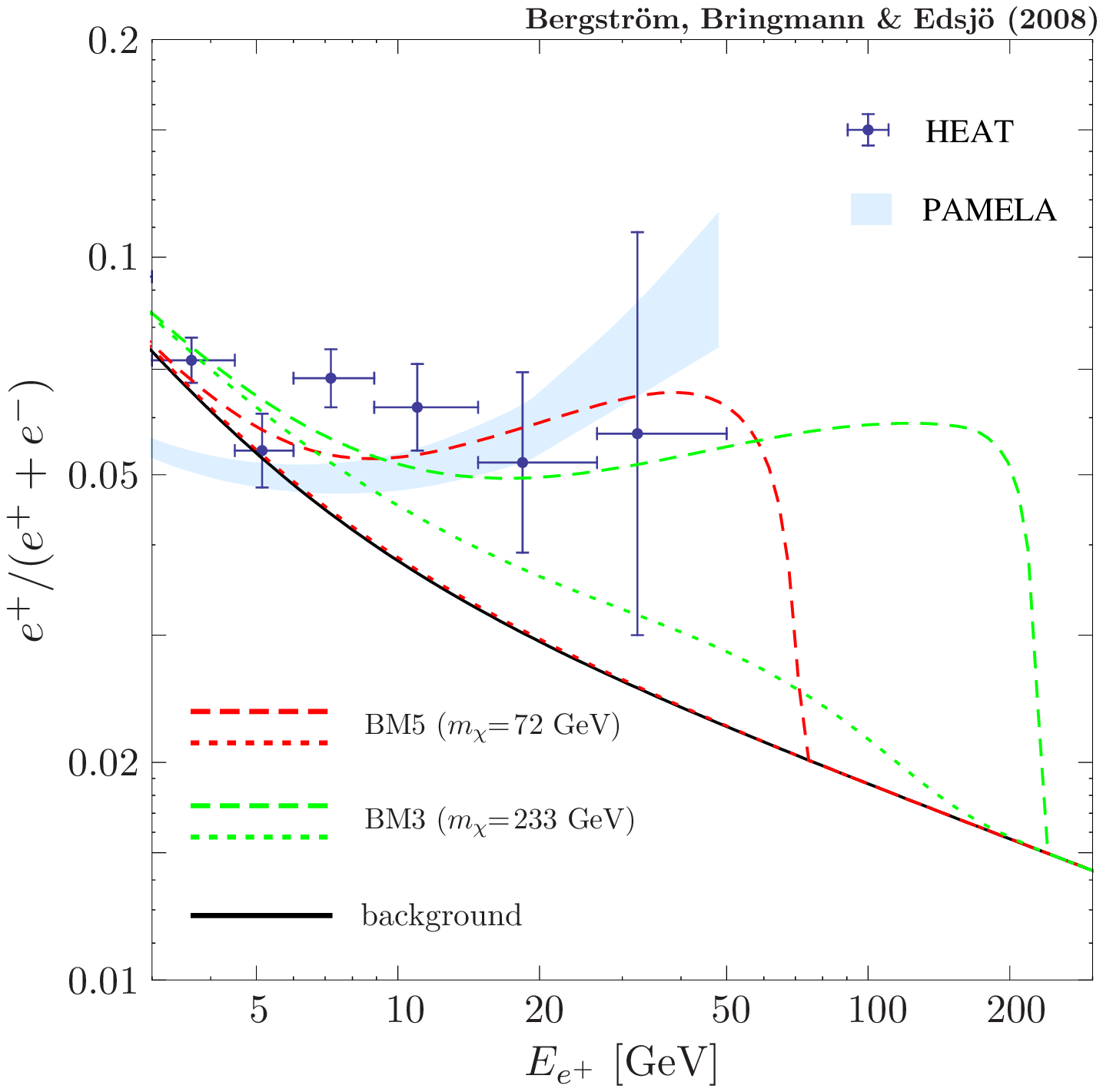}
\caption{a- (left, from~\cite{photon-photon-large}) Photon spectra expected from two particular WIMP candidates for boost factors $B= 10^4$ (red) and $B=100$ (black). The EGRET data (green), the HESS source spectrum (blue) and the FGST (old GLAST) sensitivity (dotted line) are also shown.   b- (right, from~\cite{Bergstrom:2008gr}) Preliminary positron fraction PAMELA data (light blue band) presented in this conference (not published yet), HEAT data, estimated background (black line) and  spectra expected from  two particular WIMP candidates, with (dashed lines) and without (dotted lines) radiative corrections included and$B \simeq 10^4$.}
 \label{PAMELA}
\end{figure*}

 Monochromatic photons  can be  produced through $\chi \chi \to\gamma \gamma$ (or $\gamma Z$) with energy equal (or close) to $m_{\chi}$. Detection of this monoenergetic  spectral line would be a ``smoking gun" signature. However, usually~\cite{photon-photon}, but not always~\cite{photon-photon-large} (see Fig.\ref{PAMELA}a), this processes happen only at the one-loop order and  they are suppressed with branching ratio close to  10$^{-3}$-10$^{-4}$. Secondary $\gamma$ (as well as positrons and antiprotons, also important potential signals) would be produced with a spectrum whose cutoff at high energies is $m_\chi$. Until recently the calculation of  annihilation processes which occur at tree level had not included higher order corrections. Internal bremsstrahlung processes, i.e. $\gamma$ emitted by intermediate particles, have been now shown to sometimes drastically enhance  the $\gamma$ spectrum at high energies~\cite{Int-Brem}.
  
 Gamma-ray astronomy is done with ground and space instruments.  Four large  ground Atmospheric Cherenkov Telescopes (ACTs) are operating at present:  HESS in Namivia, MAGIC in La Palma, VERITAS in the US and  CANGAROO-III in Australia. They can detect photons with energy above 20 GeV and up to several TeV.
 The GLAST  (Gamma-ray  Large Area Space Telescope) satellite
 was launched on June 11, 2008~\cite{GLAST-ICHEP08}.   After  the initial 2 months calibration period, it was renamed  Fermi Gamma-ray Space Telescope (FGST from now one). FGST will provide $\gamma$-ray spectroscopic data of unprecedented quality.  The LAT (Large Area Telescope), the main instrument on FGST, detects $\gamma$-rays in the 20 MeV to 300 GeV energy range. In the 20 MeV to 10 GeV it has more than one order of magnitude better sensitivity  than its predecessor EGRET (onboard of the Compton Gamma-ray Observatory) which observed in this range, and will have a considerable greater exposure. The smaller instrument GBM (GLAST Burst Monitor) observes in the 8 keV to 30 MeV range.
     The state of the art ``Via-Lactea" simulations conclude that FGST may well discover a few to 10's of subhaloes at the 5$\sigma$ significance level~\cite{Kuhlen:2008aw}.
 Also,  several potential WIMP signals have come up in indirect detection over the years,  most of them to be tested by the FGST~\cite{Baltz:2008wd}. 

ACTs have seen an excess of  Very High Energy (VHE) $\gamma$-rays in the 0.2 to 10 TeV from the galactic center. CANGAROO, VERITAS, HESS and MAGIC~\cite{Tsuchiya:2004wv} all reported an excess, which until 2006 could be fitted in terms of the annihilation of WIMPs with mass in the TeV range.  In  2006 HESS released new and better data  showing a power law spectrum from about 160 GeV to 20 TeV, which could accommodate at most a 10\% contribution from DM  annihilation~\cite{Aharonian:2006wh} (see Fig.~\ref{PAMELA}a). The HESS source is consistent with point like  emission coincident with the position of the supermassive black hole at the center of the Milky Way or another astrophysical source very close to it. Instead of providing a signal of DM, the $\gamma$-ray excess due this source constitutes a background that limits the capability of FGST to detect DM in the TeV energy range~\cite{Zaharijas:2006qb}. The possibility of DM detection depends now on the exact distribution of DM near the galactic center and the angular resolution of FGST (which increases with increasing energy, from a few degrees at 10 MeV to less that 0.1 degrees above 10 GeV)~\cite{Dodelson:2007gd}.

The emission of 511 keV photons from the center of the galaxy was  first observed by balloon born $\gamma$-ray detectors during the 1970's. This is a clear signal of  non-relativistic positrons annihilating with electrons almost at rest. There has been  a vigorous  debate about the origin of these positrons ever since. The isotropy of the emitting region, which seamed spherical and centered on the center of the galaxy, was one of the main reasons to consider DM annihilation as the origin of the positrons. Only the DM distribution is expected to be spherically symmetric near the galactic center. It was argued that any astrophysical origin should show in some correlation with the visible matter distribution in the region and none had been observed. This changed in January of this year. The analysis of four years of data from the ESA satellite INTEGRAL launched in 2002,  have revealed an asymmetry in the emitting region, which is more extended towards the galactic plane, and  with twice as much emission on  one side than in the other of the galactic center~\cite{Weidenspointner:2008zz}. INTEGRAL also found evidence that a population of binary stars (called low mass X-ray binaries), known potential sources of  positrons, is also off-center and corresponding in extent to the observed cloud of antimatter. These observations have decreased the motivation to consider DM annihilation or decay as the origin of the signal, although it is yet unclear if it can be explained satisfactorily solely with astrophysics.  Special DM candidates were proposed to explain this signal, since the annihilations of usual WIMPs would produced positrons with too high energies. Positrons must be produced with no more than a few MeV of energy. Thus, these  DM candidates either have masses of a few MeV (they are called LDM, Light DM~\cite{Boehm:2003bt}-see also~\cite{Sahu}), or have an excited state (to which particles would be collisionally excited) which decays to the fundamental state releasing an energy in the MeV range, although the particle mass is close to 500 GeV (these are called XDM, eXciting DM~\cite{Finkbeiner:2007kk}).

 EGRET has observed an excess in the flux of 1-10 GeV diffuse  galactic $\gamma$-rays of about a factor of 2  with respect to what is expected from current models of the galactic halo~\cite{Hunger:1997we}. This  ``EGRET excess" or ``GeV excess" is compatible with that expected from the annihilation of  50-100 GeV WIMPs in an almost spherical isothermal halo with a particular substructure in the galactic  plane~\cite{deBoer:2005tm}.
FGST will  check the ``EGRET excess" and observe  the energy region  between EGRET and HESS, from 10 to 200 GeV (see Fig.~\ref{PAMELA}a). 

The ``WMAP Haze"~\cite{Finkbeiner:2003im} is an excess of microwave emission in the inner 20 degrees, about 1 kpc,  around the galactic center, that remains once all other expected foregrounds due to all the standard emission mechanisms in the interstellar medium have been subtracted. The excess still persists in the recent 5-year data release of WMAP~\cite{dobler-private}.  It is interpreted as synchrotron emission with a very hard spectral index, consistent with synchrotron emission from highly relativistic electron-positron pairs produced by WIMP annihilation and deflected in the galactic magnetic field~\cite{Finkbeiner:2004us, Hooper:2007gi}. Supernovae shocks also produce relativistic electrons which emit synchrotron radiation, but it is argued that diffusion and energy losses would make  the spectrum of this radiation not as hard as observed. Thus the argument in favor of a DM explanation resides in the spectral index of the radiation. 
 If the DM interpretation is correct, the Haze constrains the distribution of  DM within the  inner several kpc of our galaxy, but not the mass of the  WIMP. Any WIMP with mass in the range 50 GeV to a few TeV and annihilating into fermions or gauge bosons would generate a synchrotron emission within a factor of 2 to 3 of that observed in the WMAP Haze~{Hooper:2007gi}.   WIMP annihilations explaining the Haze would also produce a flux of $\gamma$-rays within the reach of FGST~\cite{Hooper:2007gi} (X-ray and radio observations might also be important~\cite{Regis:2008ij}).

Unlike photons, positrons and antiprotons  produced in WIMP annihilations would not keep any directional information because of the effect of the magnetic fields in the disk of our galaxy (and at energies smaller than a GeV, also the effect of the solar wind close to Earth). Positrons and antiprotons, which would be produced in equal numbers as electrons and protons, are an interesting potential signal of WIMP annihilation because there is not much antimatter in the Universe. The signal would consist of an excess of $e^+$ or $\bar p$ with a characteristic spectrum  presenting a cutoff at high energies (related to the WIMP mass) with respect to the approximately 
power-law spectrum expected from secondary production in cosmic ray propagation (see Fig.~\ref{PAMELA}b).  Balloon born experiments detecting positrons have found since the 1980's a possible excess, the so called ``HEAT excess"~\cite{Barwick:1997ig}, also found by AMS, which could be explained with WIMP annihilations (with WIMP masses above 200 GeV and annihilation  boost factors $B>$ 30~\cite{Baltz:1998xv}.)

PAMELA, a satellite carrying a magnetic spectrometer launched in June 2006, has reported in this conference~\cite{Boezio} their much anticipated first positron results.  PAMELA reported an excess in the positron fraction (the ratio of $e^+$ flux and $(e^+ + e^-)$ flux) in the 10 to 60 GeV energy range compatible with the ``HEAT excess". The PAMELA data on the antiproton to proton ratio released in February 2008 is instead compatible with secondary cosmic rays. 
The preliminary PAMELA positron ratio data show six points from about 10 to 50 GeV with positron fractions raising steadily from close 0.05 at 10 GeV to 0.10 at 50 GeV (see the light blue band in Fig.~\ref{PAMELA}b taken from a paper appeared  just after this conference) while cosmic rays secondaries should have positron fractions diminishing steadily with increasing energy in the same energy range. Solar modulation effects, important at energies below 1GeV, have not yet been taken into account and should explain the large dispersion in the data from different experiments at those energies.  These or other effects may  still change the calibration  of the PAMELA data. If the collaboration confirms its finding in the near future, the quality of the data  will make this excess the first strong hint of annihilating DM (certainly, astrophysical sources, such as supernovae, should be excluded as  potential sources first). PAMELA can extend its measurement of the positron flux up to 200 GeV and a cutoff  at some energy would provide an indication of the annihilating WIMP mass (as in Fig.~\ref{PAMELA}).  Since the first papers  in 1990~\cite{Kamionkowski:1990ty} attempting to explain with annihilating DM the early indications of a positron  excess until the very recent papers already fitting the PAMELA data (appeared just after the ICHEP08 conference)~\cite{Bergstrom:2008gr, Barger:2008su}, large boost factors of the annihilation rate are required to account for the data with the usual WIMP candidates. Fig.~\ref{PAMELA}b, taken from~\cite{Bergstrom:2008gr}, shows some examples of spectra produced with
neutralinos in versions of the MSSM (Minimal Supersymmetric Standard Model), in which  $B \simeq10^4$ are assumed. These very large $B$ may be avoided with some WIMP candidates (if they are Dirac instead of Majorana fermions, for example). Otherwise the large $B$ may be explained by   the existence of larger than expected inhomogeneities in the dark halo of our galaxy or by non-standard cosmological models for the pre-BBN era (see below). Fig.~\ref{PAMELA}b also shows the importance of including radiative corrections in the calculation of the expected annihilation signal. 

Finally, if DM particle annihilations explain the positron excess confirmed by PAMELA, we expect a signal in photons, which FGST might be able to test. Thus PAMELA and FGST data will be combined with  LHC data and other collider bounds to test particle candidates~\cite{Morselli}.

 \section{DIRECT DARK MATTER SEARCHES} 
 
 There are many direct DM detection experiments running or in construction~\cite{texono}  or in the stage of research and development~\cite{Stapnes}. They use different target materials and different detection strategies. Single channel techniques measure only one of the effects produced by the recoiling nucleus  hit by a DM WIMP, usually phonons (or heat), ionization or scintillation. Among the experiments measuring only ionization (in Ge, Si or CdTe) are IGEX, HDMS, GENIUS, TEXONO and CoGeNT, among those using only scintillation (in NaI, Xe, Ar, Ne or CsI) are DAMA, NAIAD, DEAP/CLEAN, XMASS and  KIMS and among those using phonons  (in Ge, Si, Al$_2$O$_3$ or TeO$_2$) are CRESST-I, Cuoricino and CUORE. Threshold detectors search instead for bubbles produced by the energy deposited in a WIMP collision, either in a superheated bubble chamber, as done by COUPP, or  with superheated freon, C$_4$F$_{10}$, droplets suspended in a CsCl gel, as in PICASSO. Several experiments use hybrid detector techniques in which the relative intensity of two different effects is used to discriminate between nuclear recoils and the background. Experiments such as CDM, SuperCDMS, EDELWEISS and EURECA use ionization and phonons (in Ge or SI); ZEPLIN, XENON, WARP, ArDM, LUX or XAX use ionization and scintillation (in Xe, Ar or Ne); CRESST-II uses scintillation and photons
 (in CaWO$_4$). Some of these experiments, such as CDMS or EDEWEISS use also timing of the signal to provide further discrimination. CDMS, using ionization, phonons and timing, succeeds in being background fee. Liquid noble-gas detectors, using Xe, Ar or Ne, either in a single (liquid) phase and measuring scintillation, or in double (liquid-gas) phase and measuring ionization and scintillation are very promising, because their mass can in principle be easily upgraded to several tonnes~\cite{Galbiati, Stapnes}. Directional WIMP dark matter detectors, for which there are several prototypes,  are also extremely interesting~\cite{Sciolla}.
 \begin{figure*}[t]
\includegraphics[width=60mm]{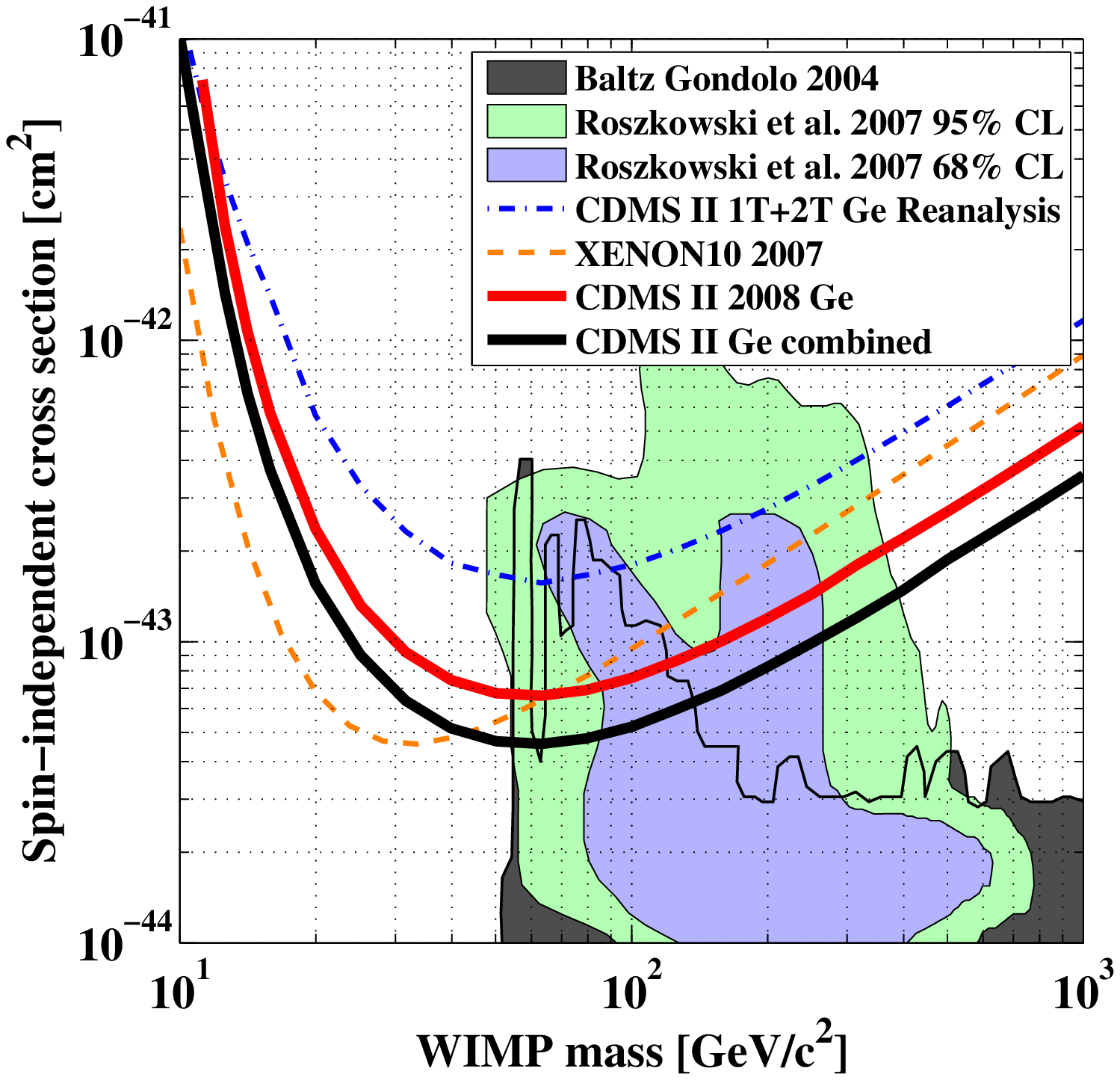}
\includegraphics[width=70mm]{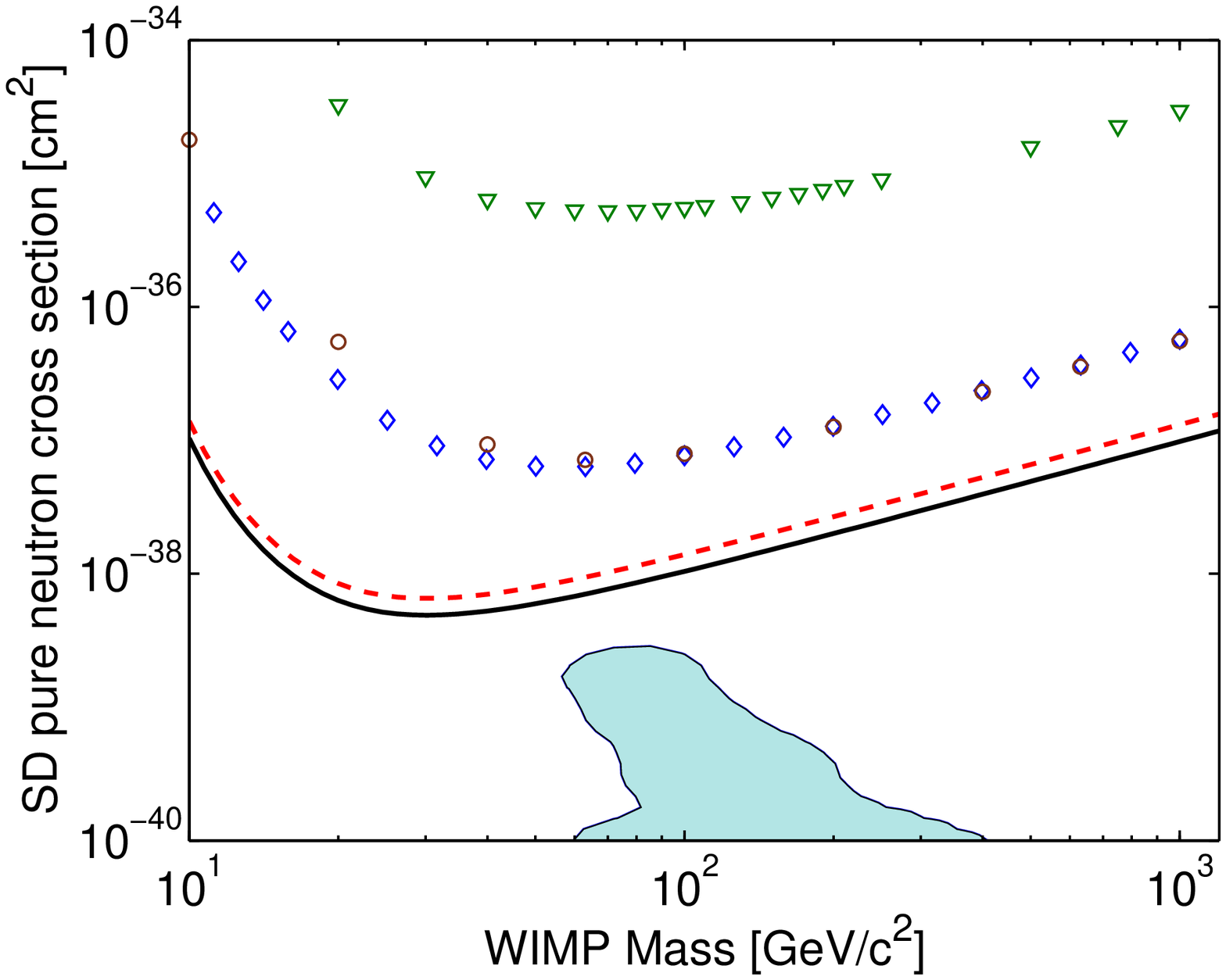}
\caption{a- (left, from~\cite{Ahmed:2008eu}) best bounds on WIMP-nucleon cross section for SI interactions. b- (right, from~\cite{Angle:2008we}) best bounds on the WIMP-neutron cross section for WIMPs with SD interaction coupled to neutrons only   from XENON-10 (solid black and dashed red lines), CDMS (diamonds), ZEPLIN-II (circles)
 and KIMS (triangles)~\cite{Akerib:2005za}. }
 \label{SI-SD-bounds}
\end{figure*}

 At present CDMS and XENON-10 compete to provide the best upper bounds on WIMPs heavier than 10 GeV (below 5 GeV,  others provide the best bounds, e.g. TEXONO and CRESST-I) with spin-independent (SI) interactions (whose cross section is enhanced by the square of the number of nucleons in the interacting nucleus) or with spin-dependent interactions (SD) (whose cross section depends on the spin of the interacting nucleus). For SI interactions CDMS provides the best bounds  on the WIMP-nucleon cross section for WIMPs heavier than 30 GeV~\cite{Ahmed:2008eu} and XENON-10 provides the best bounds for masses in the 10 to 30 GeV range~\cite{Angle:2007uj}  (see  Fig.\ref{SI-SD-bounds}a taken from~\cite{Ahmed:2008eu}). XENON-10 has the best bounds on WIMPs with SD interactions coupled only to  neutrons~\cite{Angle:2008we} (see Fig.\ref{SI-SD-bounds}b from Ref~\cite{Angle:2008we}). Super Kamiokande provides instead the best bounds on WIMPs with SD interactions which couple only to protons (at the level of 10$^{-38}$ for the WIMP-proton cross section) followed by KIMS,  COUPP and NAIAD~\cite{Desai:2004pq} 

Contrary to the negative results of all other direct DM searches,  the DAMA
collaboration has found an annual modulation in their data compatible
with the signal expected from dark matter particles bound to our
galactic halo and a standard halo model. The  annual modulation seen by the DAMA/NaI experiment~\cite{Bernabei:2003za} and confirmed by the new experiment
DAMA/LIBRA~\cite{Bernabei:2008yi} of the same collaboration, is the
only positive signal seen in any direct DM search. The 7 years of data of DAMA/NaI showed a 6$\sigma$ modulation signal.  Now, the combined 11 years of data of both experiments (with a very impressive exposure,  0.83 ton$\times$year) show an 8.2$\sigma$ annual  modulation signal.  An annual modulation is expected due to the motion of the Earth around the Sun. In a standard halo, the  average WIMP velocity relative to Earth  is maximum when the Earth velocity around the Sun adds up maximally to the velocity of the Sun around the galaxy (June 2), and is minimum six months later (December 2).  The DAMA modulation has the phase expected in a standard halo (important streams of DM passing through Earth could change the phase).

Are the DAMA results compatible with those of all other negative searches? There are many aspects to this question and  I will concentrate on WIMPs which scatter elastically with nuclei and a standard halo model. Papers written prior to the latest DAMA/LIBRA results found
regions of WIMP mass and cross section that reconciled all null
results with DAMA's positive signal, even assuming a standard halo
model: WIMPs with spin independent
interactions~\cite{Gelmini:2004gm}
in the mass range 5--9~GeV, and WIMPs with spin dependent
interactions~\cite{Savage:2004fn} in the mass range 5--13~GeV were
found to be compatible with all existing data. In addition, Ref.~\cite{Gelmini:2004gm} studied  also models with DM streams and
found that nonstandard velocity distributions could reconcile all
the existing data sets including DAMA. 

The upper bounds have now improved and the regions that were found to be compatible in 2005 are now rejected. But ion channeling had not then been taken  into account. The ion channeling effect in crystals has been studied since the 1960's (see e.g.~\cite{Drobyshevski:2007zj} and references therein) and changes the energy fraction deposited into electrons. In general only a fraction (known as the quenching factor $Q$) of the recoil energy deposited by a WIMP is transferred to electrons  and goes into the scintillation signal observed by DAMA. The rest is converted into phonons/heat as the recoiling nucleus collides with other nuclei and is not observed. The DAMA detector is composed of NaI and $Q_{Na} = 0.3$, $Q_I = 0.09$. However, nuclei that recoil
along the characteristic axes or planes of the crystal structure may
travel long distances without colliding with other nuclei. In these cases, which are more likely for small recoil energies, the measured energy corresponds  to the energy deposited by the WIMP, thus $Q \approx 1$~ \cite{Bernabei:2007hw}.  As a consequence, the detector is sensitive to lower mass WIMPs than previously thought.  The DAMA threshold is 2~keVee (electron
equivalent). Using the above quenching factors this was thought to
correspond to 7~keV and 22~keV recoil energy for Na and I respectively; yet, due to channeling the actual threshold for the channeled  events can be as low as 2~keV. 

 The new channeling studies revived the possiblity of DAMA's compatibility with other data sets, particularly at low masses.
 \begin{figure*}[t]
\rotatebox{90} {\includegraphics[width=48mm]{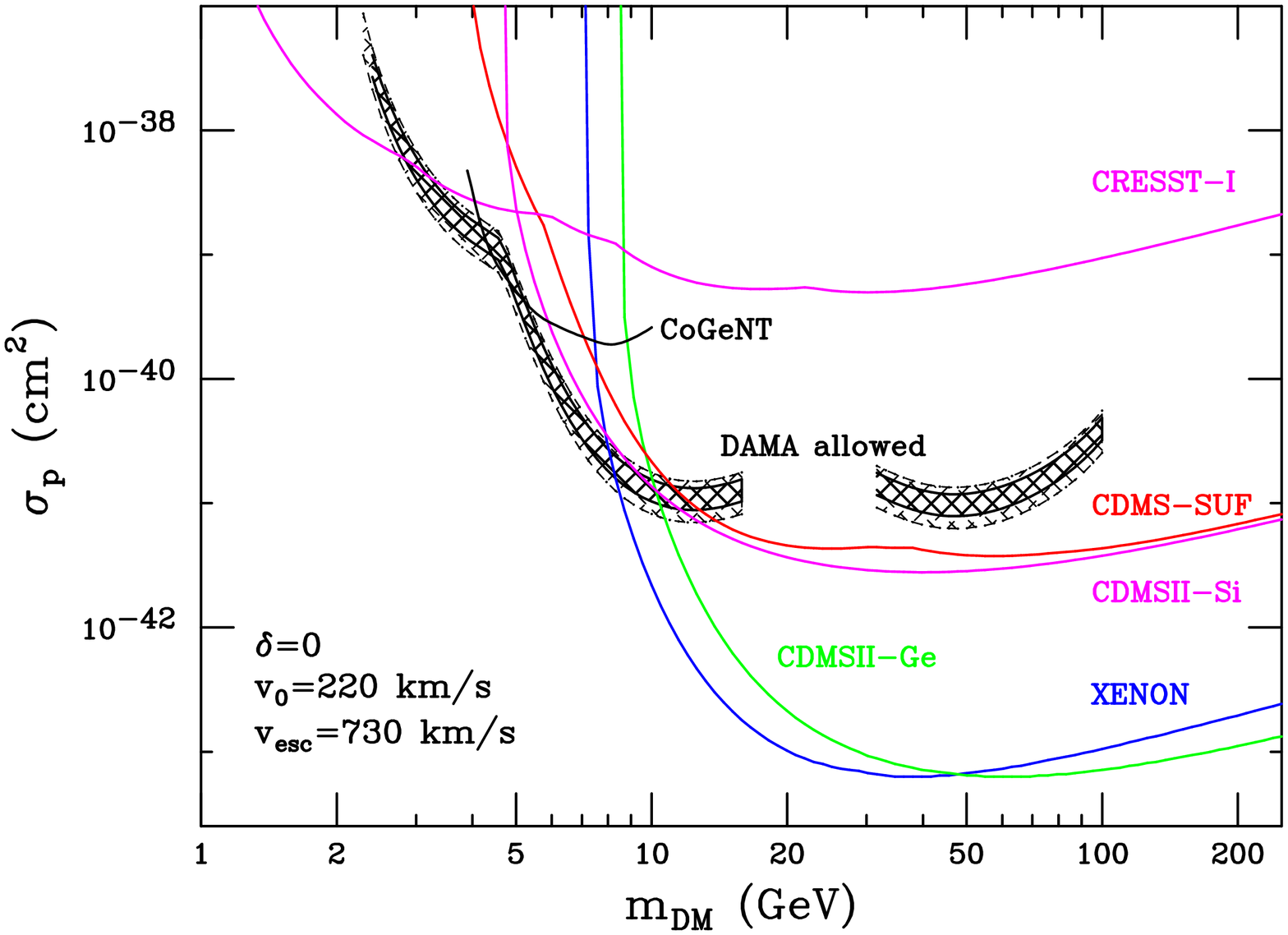}}
\includegraphics[width=102mm]{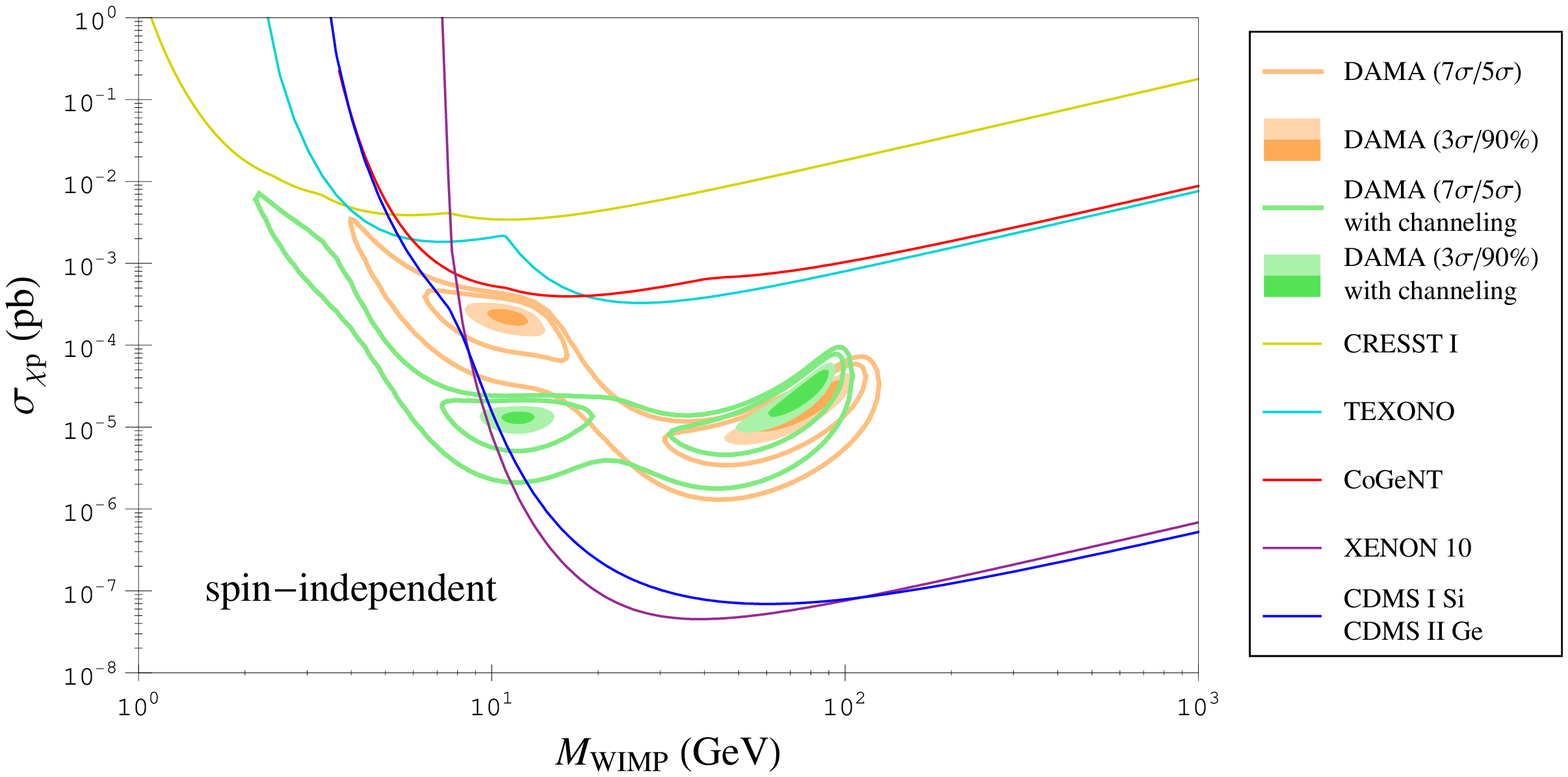}
\caption{Experimental constraints and DAMA preferred region of WIMP-proton cross section versus WIMP mass for SI only scattering. a- (left) DAMA region obtained with a raster scan in the WIMP mass and 2 modulation amplitude data bins, with channelling~\cite{Petriello:2008jj}. b- (right) DAMA regions determined with a likelihood ratio method and 36 modulation data bins, with (green) or without (orange) ion channeling included~\cite{Savage:2008er}. }
 \label{libra}
\end{figure*}
Repeating the analysis of~\cite{Gelmini:2004gm} with the new data,  Ref.~\cite{Petriello:2008jj} found  light WIMPs  with SI interactions compatible with all experimental results, when channelling is included (Fig.~\ref{libra}a).  Ref.~\cite{Petriello:2008jj}  used the DAMA modulation data  binned only into two bins. The latest DAMA/LIBRA paper~\cite{Bernabei:2008yi} provided the modulation data also in 36 bins, thus  including  spectral modulation amplitude information. This was used in all the other papers  that appeared just after the ICHEP08 conference  studying the compatibility of the DAMA
signal with other negative experimental results, assuming
the DM consist of  WIMPs with elastic interactions~\cite{Bottino:2008mf,
 Savage:2008er} (although WIMPs with inelastic interactions~\cite{Chang:2008gd} and non-WIMP candidates, such as mirror, composite, and WIMPless DM~\cite{Foot:2008nw} were studied as well). The spectral modulation amplitude information favours heavier WIMPs. For SI interactions, the most preferred DAMA regions of
any WIMP mass are ruled out to 3$\sigma$, even with channeling taken into account~\cite{Savage:2008er}. However, for WIMP masses of $\sim$10~GeV 
some parameters outside these regions still yield a reasonable fit to
the DAMA data and are compatible with all 90\% C.L.\ upper limits from negative searches, when channeling is included.~\cite{Bottino:2008mf, Savage:2008er}. For SD interactions also regions of compatibility are found for WIMP masses below 10 GeV, when channeling is included~\cite{Savage:2008er}. Channelling should also happen in all other crystals, such as Ge and Si, thus some of  the upper bounds of negative searches might be somewhat affected by channeling, specially for low mass WIMPs.

 \section{WIMP RELIC ABUNDANCE} 

The argument showing that WIMPs  are good DM candidates is old an well known. The density per comoving volume of non relativistic  particles in equilibrium in the early Universe decreases exponentially with decreasing temperature, due to the Bolzmann factor, until the reactions which change the particle number become ineffective.  
 At this point, when  the annihilation rate  becomes smaller than the Hubble expansion rate, 
  the WIMP number per comoving volume becomes constant. This  moment of chemical decoupling or freeze-out happens later, i.e. for smaller WIMP densities, for larger annihilation cross sections $\sigma$. If there is no subsequent change of entropy in matter plus radiation, the present relic density is $\Omega_{\rm std} h^2 \simeq 10^{-10} {\rm ~GeV^{-2}}/ {\left< \sigma v \right> } $, which for weak cross sections gives the right order of magnitude of the DM density (and a temperature  $T_{f.o.} \simeq m_\chi/20$ at freeze-out for a WIMP of mass $m_\chi$). 

 This is a ballpark argument. When actually applied to particle models, the requirement that the WIMP candidate of the model must have the measured DM density is very constraining. In many SUSY models, in which the WIMP candidate is usually a neutralino, this ``DM constraint"  is very effective in restricting the parameter space of models. In minimal supergravity models (mSUGRA) for instance, the neutralino typically has a  small annihilation rate in the early Universe, thus its relic density  tends to be larger than observed. The  ``DM constraint"  is found to  be satisfied only along  four very  narrow regions in  the fermionic and scalar mass parameter space $m_{1/2}$, $m_0$. Most of the ``benchmark points", special models chosen to study in detail in preparation for the  LHC  and the next possible collider,  lie on those very narrow bands (which become more fine-tuned for large $m_{1/2}$ and $m_0$ values).
  Neutralinos are underabundant (account for a fraction of the DM) also in narrow regions adjacent to these just mentioned, but  in most of the parameter space neutralinos are overabundant and the corresponding models are thought to be rejected.
Is it  correct to reject all these SUSY  models? 

The issue is that the narrow bands just mentioned  depend not only on the particle model to be tested in collider experiments, but on the history of the Universe before BBN, an epoch from which we have no data. BBN is the earliest episode (200 s  after the Bang,  $T\simeq 0.8$ MeV) from which we have a trace, the abundance of light elements
D, $^4$He and $^7$Li. The next observable is the CMB radiation (3.8 $~10^4$ yr after the Bang,  $T\simeq$ eV) and the next is the large scale structure of the Universe. WIMP's have their number fixed at $T_{f.o.} \simeq m_\chi / 20$, thus WIMPs with $m_\chi \geq 100$ MeV would be the earliest remnants ever studied and, if discovered,  they would  for the first time give  information on the pre-BBN epoch of the Universe.
As things stand now, to compute the WIMP relic density we must make assumptions about the pre-BBN epoch. The standard assumptions are reasonable, but are just that, assumptions. The standard computation of the relic density relies on the  assuming that the entropy of matter and radiation is conserved,
that WIMPs are produced thermally, i.e. via interactions with the particles in the plasma,  and were in kinetic and chemical equilibrium before they decoupled at $T_{f.o.}$. These are just assumptions, which do not hold in many cosmological models. These include models with moduli decay, Q-ball decay and thermal 
inflation~\cite{Moroi-etc}, in which there is a late episode of entropy production or inflation and  non-thermal production of the WIMPs in particle decays is possible. It is enough that the
highest temperature of the radiation dominated period in which BBN happens, the so called
reheating temperature $T_{RH}$, is larger than 4 MeV~\cite{hannestad} for BBN and all the subsequent  history of the Universe to proceed as usual.
In non-standard cosmological models the WIMP relic abundance may be decreased  or increased with respect to the standard abundance. The density may be decreased by reducing the rate of  thermal production (through a low $T_{RH} < T_{f.o.}$~\cite{Chung:1998ua}) or producing radiation after freeze-out (entropy dilution). The density may be increased by creating WIMPs from particle (or extended objects) decay (non-thermal production)~\cite{gg} or increasing the expansion rate of the Universe at freeze-out (e.g. in kinetion domination models~\cite{kinetion}). 

In models in which a  scalar field (one of the moduli fields pervasive in string or plain SUSY models or an inflaton) dominates the energy density of the Universe and decays late  producing a plasma with a low temperature $T_{RH}$,  the WIMP density depends on two  additional parameters besides the usual ones, one of them being $T_{RH}$~\cite{gg}. These parameters depend on the completion of the  models to higher energy scales than those that will be tested in colliders.  
 \begin{figure*}[t]
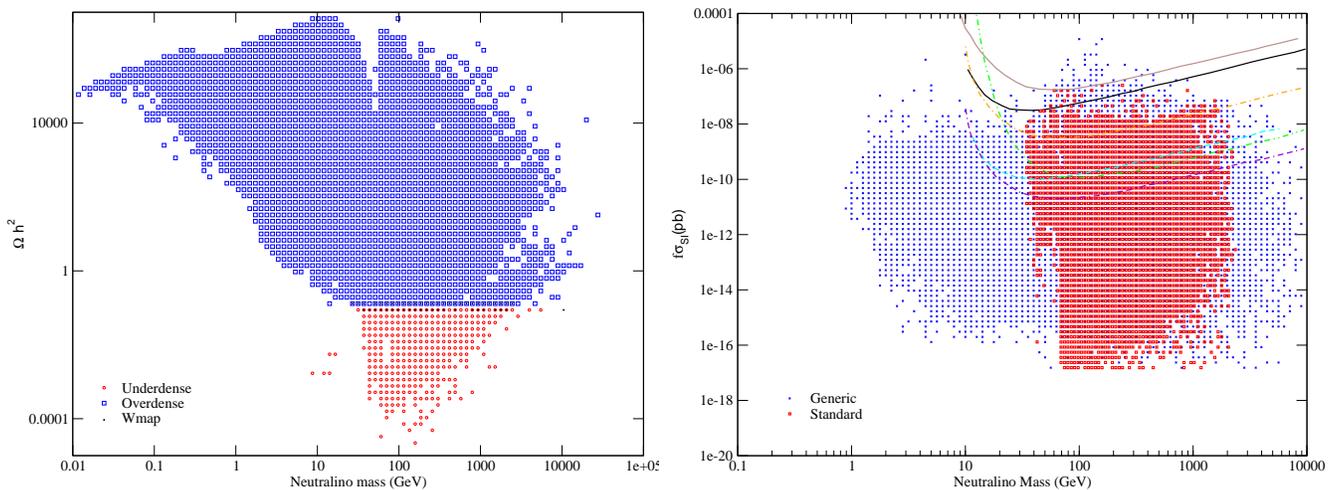

\centering
\includegraphics[width=0.49\textwidth]{omegaunov.eps}
\includegraphics[width=0.49\textwidth]{gracgrid.eps}
\vspace{-10pt}
\caption{a-(left) Standard neutralino density $\Omega h^2$ and b-(right) halo fraction times SI proton-neutralino scattering cross section, as a function of the mass, from Ref.~\cite{ggsy-2} (color indicates standard relic densities).}
\vspace{-10pt}
 \label{array}
\end{figure*}
Provided the right values for the two parameters can be obtained in the theory at the high scale, neutralinos can have the DM density in (almost) all SUSY models (the exception being severely overabundant or underabundant very light neutralinos~\cite{ggsy-2}, rarely encountered in SUSY models). This has important implications not only for colliders but for direct and indirect DM searches as well, since now the region of parameter space is much larger.
We see in Fig.~\ref{array}a the standard relic density region as function of the mass of $10^5$ MSSMs defined by 10 parameters at the electroweak scale~\cite{ggsy-2}. Here the bino mass parameter $M_1$ was allowed be much smaller than the other two gaugino mass parameters, as low as 100 MeV, which does not contradict any experimental bound (in models in which the three gaugino masses are not required to coincide at a large energy scale). All the models above the black strip showing the right DM abundance are rejected in the standard cosmology, because the neutralinos they predict are overabundant. Only the models with the right DM density or less (black-red regions) are usually assumed to be viable and their halo fraction times SI proton-neutralino cross section  $f \sigma_{SI}$ (on which the interaction rate  in direct searches depends) fall in the red-black region of Fig.~\ref{array}b, with masses  30 GeV to 2 TeV. Assuming instead that all the models whose density can be brought to coincide with the DM density in~\cite{gg} have the DM density, the region of models to be searched for in direct detection experiments changes to the blue region of Fig.~\ref{array}b, which extends from  under 1 GeV to  10 TeV in mass. Fig.~\ref{array}b shows also present (black solid line) and future expected experimental limits (see \cite{ggsy-2} for details). It is interesting that, following this line of arguments, very light neutralinos, much lighter than 1 GeV, are allowed in SUSY models  and may constitute the DM, as presented by S. Profumo in this conference~\cite{profumo}.
 
 Not only the relic density of WIMPs but their characteristic speed before structure formation in the Universe can differ in standard and non-standard pre-BBN cosmological models. In the latter WIMPs could be colder, thus leading to the formation of much smaller subhaloes and thus larger values of the annihilation boost factor $B$~\cite{Gelmini:2008sh}.
 
Assume that the LHC finds a DM candidate and determines its standard relic density  to be  much larger than the DM density, say  $10^4$. Then, either the particle is unstable (e.g. in SUSY models it is the next-to-lightest particle, the NLSP, and the SUSY spectrum should tell if this is true) or it does constitute the DM (it should be found in DM searches then) and the pre-BBN cosmology is non standard.

In conclusion, in most scenarios one can think of the LHC should find at least  a hint of  new physics, and whatever it finds will lead to a set of possible DM candidates and reject others. DM searches are independent and complementary to collider searches in multiple ways and they are advancing fast. Direct and indirect DM searches are complementary to each other. The DM may have several components to be found in different ways.  If discovered WIMPs
would be our first probe of the cosmology immediately before BBN. All possibilities are still open, but hopefully not for long.
 
\begin{acknowledgments}
This work was supported in part by DOE grant DE-FG03-91ER40662 Task C.
\end{acknowledgments}


\begin{thebibliography}{9}  

\bibitem{bullet-cluster} 
  D.~Clowe, S.~W.~Randall and M.~Markevitch,
  Nucl.\ Phys.\ Proc.\ Suppl.\  {\bf 173}, 28 (2007).
 
 \bibitem{WMAP-5year} 
  E.~Komatsu {\it et al.}  [WMAP Coll.],
  arXiv:0803.0547 [astro-ph].

  \bibitem{MACHOs} 
  J.~Yoo, J.~Chaname and A.~Gould,
  Astrophys.\ J.\  {\bf 601}, 311 (2004);
  P.~Tisserand {\it et al.}  [EROS-2 Coll.],
  Astron.\ Astrophys.\  {\bf 469}, 387 (2007);
  J.~Rich  [EROS-2 Coll.],
  Nucl.\ Phys.\ Proc.\ Suppl.\  {\bf 173}, 40 (2007).

 \bibitem{LHC-DM-ICHEP08} See e.g. the contribution of David Toback to this conference.

\bibitem{Rott} See e.g. the contribution of Carsten Rott to this conference.

\bibitem{deJong} See e.g. the contribution of  Maarten de Jong to this conference.

\bibitem{GLAST-ICHEP08} See the contribution of Richard Dubois to this conference.

  \bibitem{Boezio} See the contribution  of Mirko Boezio to this conference.

\bibitem{Romana} See e.g. the contribution of  Francesca Romana to this conference.

 \bibitem{subhaloes} See e.g. V.~Berezinsky, V.~Dokuchaev and Y.~Eroshenko,  Phys.\ Rev.\  D {\bf 77}, 083519 (2008)
and references therein.

\bibitem{photon-photon}
  L.~Bergstrom and H.~Snellman,
  Phys.\ Rev.\  D {\bf 37}, 3737 (1988);
  L.~Bergstrom and P.~Ullio,
  Nucl.\ Phys.\  B {\bf 504}, 27 (1997);
  Z.~Bern, P.~Gondolo and M.~Perelstein,
  Phys.\ Lett.\  B {\bf 411}, 86 (1997);
  P.~Ullio and L.~Bergstrom,
  Phys.\ Rev.\  D {\bf 57}, 1962 (1998).

\bibitem{photon-photon-large} 
  M.~Gustafsson, E.~Lundstrom, L.~Bergstrom and J.~Edsjo,
  Phys.\ Rev.\ Lett.\  {\bf 99}, 041301 (2007).

\bibitem{Int-Brem} 
  T.~Bringmann, L.~Bergstrom and J.~Edsjo,
  JHEP {\bf 0801}, 049 (2008).

\bibitem{Kuhlen:2008aw}
  M.~Kuhlen, J.~Diemand and P.~Madau,
  arXiv:0805.4416 [astro-ph].  

\bibitem{Baltz:2008wd}
  E.~A.~Baltz {\it et al.},
  JCAP {\bf 0807}, 013 (2008).
  
\bibitem{Tsuchiya:2004wv}
  K.~Tsuchiya {\it et al.}  [CANGAROO-II Coll.]
   Astrophys.\ J.\ {\bf 606}, L115 (2004);
  K.~Kosack {\it et al.}  [VERITAS Coll.]
  Astrophys.\ J.\ {\bf 608}, L97 (2004);
  F.~Aharonian {\it et al.}  [HESS  Coll.]
  Astron.\ Astrophys.\ {\bf 425}, L13 (2004);
  J.~Albert {\it et al.}  [MAGIC  Coll.]
  Astrophys.\ J.\ {\bf 638}, L101 (2006).

\bibitem{Aharonian:2006wh}
  F.~Aharonian {\it et al.}  [HESS Coll.],
  Phys.\ Rev.\ Lett.\  {\bf 97}, 221102 (2006)
  [Erratum-ibid.\  {\bf 97}, 249901 (2006)].

\bibitem{Zaharijas:2006qb}
  G.~Zaharijas and D.~Hooper,
  Phys.\ Rev.\  D {\bf 73}, 103501 (2006).
  
\bibitem{Dodelson:2007gd}
  S.~Dodelson, D.~Hooper and P.~D.~Serpico,
  Phys.\ Rev.\  D {\bf 77}, 063512 (2008).
   
\bibitem{Weidenspointner:2008zz}
  G.~Weidenspointner {\it et al.},
  Nature {\bf 451}, 159 (2008).
 
\bibitem{Boehm:2003bt}
  C.~Boehm  {\it et al.}
  Phys.\ Rev.\ Lett.{\bf 92}, 101301(2004);
  D.~Hooper and K.~Zurek
  Phys.\ Rev.\  D{\bf 77}, 087302(2008).
  
  \bibitem{Sahu} See e.g. the contribution of Narendra Sahu to this conference.
  
\bibitem{Finkbeiner:2007kk}
  D.~P.~Finkbeiner and N.~Weiner,
  Phys.\ Rev.\  D {\bf 76}, 083519 (2007).

\bibitem{Hunger:1997we}
  S.~D.~Hunter {\it et al.},
  Astrophys.\ J.\  {\bf 481}, 205 (1997);
  I.~V.~Moskalenko, A.~W.~Strong and O.~Reimer,
  Astron.\ Astrophys.\  {\bf 338}, L75 (1998);
  A.~W.~Strong, I.~V.~Moskalenko and O.~Reimer,
  Astrophys.\ J.\  {\bf 537}, 763 (2000)
  [Erratum-ibid.\  {\bf 541}, 1109 (2000)];
and
  Astrophys.\ J.\  {\bf 613}, 962 (2004);
  I.~V.~Moskalenko {\it et al.}
  Nucl.\ Phys.\ Proc.\ Suppl.\  {\bf 173}, 44 (2007).
   
\bibitem{deBoer:2005tm}
  W.~de Boer {\it et al.}
  Astron.\ Astrophys.\  {\bf 444}, 51 (2005), and references therein.
  
\bibitem{Finkbeiner:2003im}
  D.~P.~Finkbeiner,
  Astrophys.\ J.\  {\bf 614}, 186 (2004);
  G.~Dobler and D.~P.~Finkbeiner,
  Astrophys.\ J.\  {\bf 680}, 1222 (2008);
  M.~Bottino, A.~J.~Banday and D.~Maino,
  arXiv:0807.1865 [astro-ph].
  
 \bibitem{dobler-private} G,~Dobler, private communication. 

\bibitem{Finkbeiner:2004us}
  D.~P.~Finkbeiner,
  arXiv:astro-ph/0409027;
  D.~Hooper, D.~P.~Finkbeiner and G.~Dobler,
  Phys.\ Rev.\  D {\bf 76}, 083012 (2007);
  G.~Caceres and D.~Hooper,
  arXiv:0808.0508 [hep-ph].
  
\bibitem{Hooper:2007gi}
  D.~Hooper {\it et al.}
  Phys.\ Rev.\  D {\bf 77}, 043511 (2008).
  
\bibitem{Regis:2008ij}
  M.~Regis and P.~Ullio,
  Phys.\ Rev.\  D {\bf 78}, 043505 (2008).

\bibitem{Barwick:1997ig}
  S.~W.~Barwick {\it et al.}  [HEAT Collaboration],
  Astrophys.\ J.\  {\bf 482}, L191 (1997).
 
\bibitem{Baltz:1998xv}
  E.~A.~Baltz and J.~Edsjo,
  Phys.\ Rev.\  D {\bf 59}, 023511 (1999);
  E.~A.~Baltz  {\it et al.}
  Phys.\ Rev.\  D {\bf 65}, 063511 (2002).
  
\bibitem{Kamionkowski:1990ty}
  M.~Kamionkowski and M.~S.~Turner,
  Phys.\ Rev.\  D {\bf 43}, 1774 (1991).
 
\bibitem{Bergstrom:2008gr}
  L.~Bergstrom, T.~Bringmann and J.~Edsjo,
  arXiv:0808.3725 [astro-ph].
  
\bibitem{Barger:2008su}
  V.~Barger  {\it et al.}
  arXiv:0809.0162 [hep-ph];
  M.~Cirelli  {\it et al.}
  arXiv:0809.2409 [hep-ph];
  N.~Arkani-Hamed  {\it et al.}
  arXiv:0810.0713 [hep-ph].
  
  \bibitem{Morselli} See the contribution of Aldo Morselli to this conference.
    
 \bibitem{texono} See the contributions of  Shin-Ted Lin (TEXONO),   Michael Akashi-Ronquest (DEAP/CLEAN), Jonghee Yoo (CDMS) and Mani Tripathi (LUX) to this conference.
 
   \bibitem{Stapnes} See also the contribution of Steinar Stapnes to this conference.
 
\bibitem{Galbiati} See e.g. the contribution of Cristiano Galbiati to this conference.

\bibitem{Sciolla} See e.g. the contribution of Gabriella Sciolla to this conference.

\bibitem{Ahmed:2008eu}
  Z.~Ahmed {\it et al.}  [CDMS Coll.],
  arXiv:0802.3530 [astro-ph].
  
\bibitem{Angle:2007uj}
  J.~Angle {\it et al.}  [XENON Coll.],
  Phys.\ Rev.\ Lett.\  {\bf 100}, 021303 (2008).
   
\bibitem{Angle:2008we}
  J.~Angle {\it et al.},
  arXiv:0805.2939 [astro-ph].
    
\bibitem{Akerib:2005za}
  D.~S.~Akerib {\it et al.}  [CDMS Coll.],
  Phys.\ Rev.\  D {\bf 73}, 011102 (2006);
  G.~J.~Alner~al.  [ZEPLIN-II Coll.],
  Phys.\ Lett.\  B {\bf 653}, 161 (2007);
  H.~S.~Lee. {\it et al.}  [KIMS Coll.],
  Phys.\ Rev.\ Lett.\  {\bf 99} (2007) 091301.
 
\bibitem{Desai:2004pq}
  S.~Desai {\it et al.}  [Super-K Coll.],
  Phys.\ Rev.\  D {\bf 70}, 083523 (2004)
  [Erratum-ibid.\  D {\bf 70}, 109901 (2004)].
  E.~Behnke {\it et al.}  [COUPP Coll.],
  Science {\bf 319}, 933 (2008).
  G.~J.~Alner {\it et al.}  [UKDM Coll.],
  Phys.\ Lett.\  B {\bf 616}, 17 (2005).
  
\bibitem{Bernabei:2003za}
  R.~Bernabei {\it et al.},
  Riv.\ Nuovo Cim.\  {\bf 26N1}, 1 (2003).
 
\bibitem{Bernabei:2008yi}
  R.~Bernabei {\it et al.}  [DAMA Collaboration],
  Eur.\ Phys.\ J.\  C {\bf 56}, 333 (2008).
 
  
\bibitem{Gelmini:2004gm}
  G.~Gelmini and P.~Gondolo,
  arXiv:hep-ph/0405278;
  P.~Gondolo and G.~Gelmini,
  Phys.\ Rev.\  D {\bf 71}, 123520 (2005).

\bibitem{Savage:2004fn}
  C.~Savage, P.~Gondolo and K.~Freese,
  Phys.\ Rev.\  D {\bf 70}, 123513 (2004).
  
\bibitem{Drobyshevski:2007zj}
  E.~M.~Drobyshevski,
  arXiv:0706.3095 [physics.ins-det].
 

\bibitem{Bernabei:2007hw}
  R.~Bernabei {\it et al.},
  Eur.\ Phys.\ J.\  C {\bf 53}, 205 (2008).
 
\bibitem{Petriello:2008jj}
  F.~Petriello and K.~M.~Zurek,
  JHEP {\bf 0809}, 047 (2008).
  
  \bibitem{Bottino:2008mf}
  A.~Bottino {\it et al.}
  arXiv:0806.4099 [hep-ph];
  S.~Chang, A.~Pierce and N.~Weiner,
  arXiv:0808.0196 [hep-ph];
  M.~Fairbairn and T.~Schwetz,
  arXiv:0808.0704 [hep-ph].
  D.~Hooper  {\it et al.}
  arXiv:0808.2464 [hep-ph].
  
\bibitem{Savage:2008er}
  C.~Savage, G.~Gelmini, P.~Gondolo and K.~Freese,
  arXiv:0808.3607 [astro-ph].
  
 \bibitem{Chang:2008gd}
  S.~Chang  {\it et al.}
  arXiv:0807.2250 [hep-ph];
  
  
\bibitem{Foot:2008nw}
  R.~Foot,
  Phys.\ Rev.\  D {\bf 78}, 043529 (2008);
  M.~Y.~Khlopov and C.~Kouvaris,
  arXiv:0806.1191 [astro-ph];
  J.~L.~Feng, J.~Kumar and L.~E.~Strigari,
  arXiv:0806.3746 [hep-ph].

\bibitem{Moroi-etc} T. Moroi and L. Randall, {\it Nucl.\ Phys.}\ {\bf
B570}, 455 (2000); M. Fujii, K. Hamaguchi, {\it Phys.\ Rev.}\ D {\bf 66},
083501 (2002); D. H. Lyth, E.D. Stewart,
{\it Phys.\ Rev.}\ D {\bf 53}, 1784 (1996).

\bibitem{hannestad} 
  M. Kawasaki, K. Kohri, and N. Sugiyama, {\it Phys.\ Rev.\ Lett.}\ {\bf 82}, 4168 (1999); {\it Phys.\ Rev.}\ D {\bf 62}, 023506 (2000); S.~Hannestad, {\it Phys.\ Rev.}\ D {\bf 70}, 043506 (2004).

\bibitem{Chung:1998ua}
  D.~J.~H.~Chung, E.~W.~Kolb and A.~Riotto,
  Phys.\ Rev.\ Lett.\  {\bf 81}, 4048 (1998);
  G.~F.~Giudice, E.~W.~Kolb and A.~Riotto,
  Phys.\ Rev.\  D {\bf 64}, 023508 (2001).
    
\bibitem{gg} G.~Gelmini, P.~Gondolo,
 {\it  Phys.\ Rev.}\  D{\bf 74}, 023510 (2006);
G.~Gelmini {\it et al.}
  {\it Phys.\ Rev.}\  D{\bf 74}, 083514 (2006).

\bibitem{kinetion} P.~Salati,
  {\it Phys.\ Lett.}\ B {\bf 571}, 121 (2003):
  S.~Profumo and P.~Ullio,
 {\it  JCAP} {\bf 0311}, 006 (2003).

\bibitem{ggsy-2} G.~Gelmini, P.~Gondolo, A.~Soldatenko and C.~Yaguna,
  {\it Phys.\ Rev.}\  D {\bf 76}, 015010 (2007).

\bibitem{profumo} See the contribution of Stefano Profumo to this conference and 
  Phys.\ Rev.\  D {\bf 78}, 023507 (2008).

\bibitem{Gelmini:2008sh}
  G.~B.~Gelmini and P.~Gondolo,
 {\it  JCAP} {\bf 10}, 002 (2008).
 
 \end{thebibliography}
\end{document}